# Searching for universal behaviour in superheated droplet detector with effective recoil nuclei

Mala Das* and Susnata Seth

*Astroparticle Physics & Cosmology Division,*
*Saha Institute of Nuclear Physics, 1/AF Bidhan Nagar, Kolkata – 700064, INDIA*

Abstract

Energy calibration of superheated droplet detector is discussed in terms of the effective recoil nucleus threshold energy and the reduced superheat. This provides a universal energy calibration curve valid for different liquids used in this type of detector. Two widely used liquids, R114 and $C_4F_{10}$, one for neutron detection and the other for WIMPs dark matter search experiment, have been compared. Liquid having recoil nuclei with larger values of LET provides better neutron – γ discrimination. Gamma (γ) response of $C_4F_{10}$ has also been studied and the results are discussed. Behaviour of nucleation parameter with the effective recoil nucleus threshold energy and the reduced superheat have been explored.

*Corresponding author, E-mail: mala.das@saha.ac.in,

 Tel : 91 33 2337 5346, Extn : 3403, Fax : 91 33 2337 4637

## 1. Introduction

Superheated droplet detector is used both for the neutron detection and the cold dark matter (WIMPs ; Weakly Interacting Massive Particles) search experiments under different operating temperatures and pressures [1-5]. Since its development in 1979, it has been mainly used in neutron dosimetry [6,7]. The detector consists of the droplets of superheated liquids of low boiling points, suspended in viscous aquasonic gel or in polymer medium [1,2]. For neutron detection, the type of the detector liquid normally used are R12 ($CCl_2F_2$ ; b.p. -29.8$^o$C), R114 ($C_2Cl_2F_4$ ; b.p. 3.7$^o$C), R134a ($C_2H_2F_4$ ; b.p. -26$^o$C) etc. $C_2ClF_5$, $CF_3I$, $C_4F_{10}$ etc are the liquids used for the dark matter search experiments [3-5]. WIMP induced nuclear recoils are similar to neutron induced nuclear recoils and therefore have several identical techniques in detection. By varying the operating condition, using liquids of different boiling points and compositions, and employing different pulse analysis techniques, neutrons, α-particles, γ – rays etc, the unwanted backgrounds for cold dark matter search experiment, can be identified. Extensive studies have been made with R114 as the sensitive liquid for the response of neutrons and γ - rays of this type of detector [8-10]. $C_4F_{10}$ is the most favoured sensitive liquid for cold dark matter search experiment for PICASSO (Project In CAnada to Search for Supersymmetric Objects) collaboration [11,12]. In the present work, we perform a comparative study between two different liquids, R114 and $C_4F_{10}$ with respect to neutron detection in a mixed neutron – γ radiation field.

In the present work, the detector was fabricated in viscous aquasonic gel medium with $C_4F_{10}$ as the sensitive liquid. In our earlier publication [10], the discrimination of neutron induced events and γ induced events for R114 detector has been described. In this work, we would like to study the similar effects in $C_4F_{10}$ detector and compare it with the observations of R114. In order to explore the neutron response, the γ response of $C_4F_{10}$ detector has been studied. The energy released during the bubble nucleation process for γ induced events is observed to be independent of operating temperature which provides important information on the γ background rejection technique. An attempt has been made successfully to determine a universal energy calibration curve with the 'effective recoil nucleus threshold energy' and the 'reduced superheat' [8] valid for different types of superheated liquids. One of the important parameters in bubble nucleation for this type of detector is the nucleation parameter [13]. It is reported that the nucleation parameter shows single line behaviour for $C_4F_{10}$ liquid with operating temperature [14]. In the present work, studies on nucleation parameter obtained from different experiments have also been carried out in order to search for the universal behaviour valid for both the liquids. The manuscript is organized by a detailed description of the experiment performed, results obtained from the experiment and discussions of the physics importance of the results, followed by a conclusion section.

## 2. Present Work

Superheated droplets of R114 and $C_4F_{10}$ have been fabricated using aquasonic gel and glycerol as the supporting medium. The droplets were produced by adding the refrigerant liquid on the top of degassed aquasonic gel in a stainless steel high pressure chamber and by stirring the liquid by a stirrer rotated with variable speed motor. The fabrication procedure is similar as described in the literature by B Roy *et. al.* [15].

The responses of $C_4F_{10}$ detector in presence of $^{137}Cs$ (32.5mCi) γ - source was measured by varying the temperature of the detector in the range of $30^oC$ to $60^oC$ at the ambient pressure of 1 atmosphere. The measurement at various temperatures has been done in order to find out the γ sensitive temperature for $C_4F_{10}$. Once the γ sensitive temperature is identified, the measurement is done for neutrons below the γ sensitive temperature. In this way, the measurement for neutrons becomes free from the γ induced events. The γ response of R114 detector has already been studied for which the fabrication process was identical as that of $C_4F_{10}$ in the present case [9,10]. The volume of both kinds of detectors was 40 ml in a glass vial with aquasonic gel as the supporting matrix of the superheated droplets. The detector was placed inside the water bath and the temperature of the bath was controlled by a temperature controller (Metravi, DTC-200) with precision $\pm 1^oC$. The acoustic pulse formed due to bubble nucleation was measured by a condenser microphone with a frequency response expanding up to about 30kHz and the traces of the electrical pulse output was recorded in digital storage oscilloscope (Agilent Technologies, MSO7032A; 350MHz, 2GSa/s). The commercially available condenser microphone contains two parallel plates in which one is made of a very light material and acts as a diaphragm. The diaphragm vibrates when struck by sound waves that change the distance between the two plates and as a result, its capacitance. This gives a corresponding electric signal and the intensity of the electric signal depends on the intensity of the acoustic signal falling on it. The development of the electronic circuit with the condenser microphone is described elsewhere [16]. The low frequency signal is associated with the final stage of bubble nucleation. After complete phase transition of a droplet, a freely oscillating vapour bubble is formed which oscillates around its equilibrium radius and the ambient equilibrium pressure with a resonant frequency (e.g. 6 kHz for $C_4F_{10}$) that depends on the density of the surrounding medium [14].

For neutrons, the recoil nuclei are produced due to interaction of neutrons with the detector liquid, and these recoil nuclei deposit their energy in the liquid according to their LET (linear energy transfer) in the medium. The energy deposition in the sensitive liquid was calculated using SRIM code [17] for the carbon, fluorine ions in $C_4F_{10}$ and carbon, chlorine, fluorine ions in R114 for the neutron of maximum energy 6 MeV. The relation between the energy of the neutron and the operating temperature of the liquid can be obtained from the expression given below [9].

$$\frac{W}{k\,r_c}(T,P) = \frac{dE}{dx}(E_n) \quad , \tag{1}$$

Where $W$ is the critical energy for bubble formation obtained from reversible thermodynamics [18], $r_c$ is the critical bubble radius and $k$ is the nucleation parameter which has been found earlier to be equal to 0.11 for neutron induced nucleation upto around 10 MeV for R114 as the sensitive liquid [9]. For $C_4F_{10}$, $k$ has been calculated at 25°C using the above expression and found to be equal to 0.15. In the calculation, the threshold neutron energy has been considered as 371 keV at 25°C obtained from experiment [19].

The present experiment was performed both with R114 and $C_4F_{10}$ in presence of $^{252}$Cf (3.2 µCi) fission neutron source at the neutron sensitive temperatures, 55°C for R114 and 35°C for $C_4F_{10}$. The experiment has also been done at both the neutron and γ sensitive temperatures, 70°C for R114 [9] and 55°C for $C_4F_{10}$.

Trace of each pulse is recorded in digitised storage oscilloscope and the power ($P$) is calculated from the amplitudes of the pulses. The power, $P$ which is a measure of the energy released during the bubble nucleation process, is defined as $\mathrm{Log}_{10}\left(\sum_i |V_i|^2\right)$ where $V_i$ is the pulse amplitude in volt of the digitized acoustic pulse at the $i^{th}$ time bin, and the summation extends over the duration of the signal [10].

For construction of the universal energy calibration curve, the 'effective recoil nucleus' has been determined for different liquids from the available experimental results. It is known than the effective recoil nucleus is different for different liquids and also depends on the energy region of interest [20]. The recoil nucleus having maximum $\frac{dE}{dx}$ at a given energy is considered as the effective recoil nucleus. The results from the neutron calibration at different temperatures for the different liquids, $C_4F_{10}$, $C_4F_8$ (b.p. -9.1°C), R114, R142B ($C_2H_3ClF_2$ ; b.p. -6.9°C) is obtained from the different experiments as reported in literatures [8,9,13,19,21]. The recoil nuclei receive the maximum energy from the incident neutron through the elastic head on collision and that energy is considered in calculating the effective recoil nucleus threshold energy. The 'effective recoil nucleus threshold energy' is then plotted for the liquids $C_4F_{10}$, $C_4F_8$, R114, R142B with the 'reduced superheat, (s)'. The reduced superheat, $s$ is expressed as, $s = \frac{T - T_b}{T_c - T_b}$ [8] where, $T_b$ is the boiling point, $T_c$ is the critical temperature and $T$ is the ambient temperature of the liquid.

Attempt has also been made to search for the single line behaviour of the 'nucleation parameter' for the two liquids, $C_4F_{10}$ and R114. There are different ways of expressing the nucleation parameter, the details are found in the literature [13]. The nucleation parameter, '$a$' and '$k$' are defined as

$$E_c = a\, r_c \overline{\left(\frac{dE}{dx}\right)}_{L_{eff}}, \text{ where } E_c = \frac{4\pi}{3} R_c^3 H + 4\pi R_c^2 \left[\gamma(T) - T\frac{d\gamma}{dT}\right] + \frac{4\pi}{3} R_c^3 P_o$$

And,

$$W = k\, r_c \left(\overline{\frac{dE}{dx}}\right)_{L_{eff}}, \text{ where } W = \frac{16\pi \gamma^3(T)}{3[P_v(T) - P_o]^2}$$

Here $E_c$ = critical energy required for bubble formation, $H$ = evaporation heat per unit liquid volume, $R_c$ = critical radius of the vapour bubble, $\gamma(T)$ = surface tension at the temperature T, $P_v(T)$ = vapour pressure of the liquid, $P_o(T)$ = ambient liquid pressure and $L_{eff}$ = effective path length along the track of the particle [13]. The energy deposition over the length, $L_{eff}$ plays a significant role in bubble nucleation. Available results on nucleation parameters from different experiments have been expressed in terms of the parameter '*a*' and also calculated from Eq. (1), by converting '*k*' to '*a*' using the following expression [13]

$$a = k \frac{E_c}{W} \qquad (2)$$

3. Results and Discussions

The result of the measurements with $^{137}$Cs γ-source as a function of temperature is shown in Fig.1. In the present experiment, the threshold temperature for γ sensitivity of the unpurified detector fabricated in a normal (non-cleanroom) condition is found to be in the range of 45°C< $T_{th}$ ≤ 50°C. Therefore, studying the neutron sensitivity in the present work, we restrict to temperatures well below 45°C in the γ insensitive region. We note in this connection that the γ sensitive threshold temperature of purified $C_4F_{10}$ detector, fabricated in clean room environment and irradiated with $^{22}$Na (0.7µCi), as reported by PICASSO experiment is above 55°C [11,12]. Differential power (*P*) distribution [10] of the pulses obtained at different temperatures for $^{137}$Cs γ-rays induced events in $C_4F_{10}$ is shown in Fig.2. It is reported that the peak of the distribution of the integrated signal power, $P_{var}$, for neutron induced events for $C_4F_{10}$ detector depends on the temperature [11]. The measured temperature dependence of the peaks serves to define temperature dependent cut on $P_{var}$ in order to reject the non-particle induced events [11,12]. Fig.2 shows that the peak of the power distribution for γ induced events is independent of the operating temperature. The present result shows that a temperature independent cut on the power distribution can be used to reject γ induced events for superheated droplet detector employing $C_4F_{10}$ as sensitive liquid in soft aquasonic gel supporting medium, for the neutron detection in mixed neutron – γ radiation field, and also for WIMPs search experiment.

The calculated values of the critical radius and the critical energy for bubble formation at 25°C for both the liquids are shown in Table-1. It is observed from Table-1 that the critical energy for bubble formation is less for $C_4F_{10}$ than that of R114. Therefore, it is expected that the incident neutron energy needed to trigger the nucleation would be less for $C_4F_{10}$ than for R114, which is also observed in experiments [8,9,13,19], as shown in Fig.3. For lower energy neutron detection, $C_4F_{10}$ would be the good candidate as sensitive liquids in superheated droplet detector. The solid line in Fig.3 is obtained by employing Eq.(1) and the experimental points are obtained from different monoenergetic neutron experiments with $C_4F_{10}$ and R114 by different workers [8,9,13,19].

The $\frac{dE}{dx}$ curves of different constituent nuclei in R114 and $C_4F_{10}$ are shown in Fig.4. The two horizontal lines (dashed and dashed-dotted) in Fig.4 are the critical LETs for bubble nucleation in R114 at 55°C and in $C_4F_{10}$ at 35°C, calculated using Eq.(1). The Fig.4 shows that $\frac{dE}{dx}$ of chlorine is much higher in the lower energy region than that of other recoil nuclei, namely carbon (C) and fluorine (F) in both the liquids. Since the peak energy of neutrons from $^{252}$Cf fission neutron source occurs at about 700 keV [16], the maximum energies of recoil nuclei obtained from elastic head on collision with neutron are 74.6 keV, 198.8 keV, 133.0 keV for chlorine, carbon and fluorine respectively. At these low recoil energy region, because of the high LET of chlorine, energy deposited within the effective length required for bubble nucleation at a given temperature is larger in R114 than those due to carbon and fluorine in $C_4F_{10}$. Therefore, the discrimination of events caused by nuclear recoils and by the γ - rays are more prominent in R114 than that in $C_4F_{10}$. In Fig.5, $C_4F_{10}$ at 35°C and R114 at 55°C are sensitive to neutrons from $^{252}$Cf source and $C_4F_{10}$ at 55°C and R114 at 70°C become also sensitive to γ - rays from $^{252}$Cf source. Fig.5 shows that for R114, neutron and γ induced events are well separated, while for $C_4F_{10}$ it is not. It is also clear from Fig.5 that the *P*-distribution for γ induced events for both the liquids, $C_4F_{10}$ and R114, falls in the same region and it does not depend on the composition of the liquid.

In order to find out a universal energy calibration curve, the data are plotted in Fig.6, in terms of the 'threshold neutron energy' for the liquids R114, $C_4F_{10}$, $C_4F_8$, R142B obtained from different experiments with the 'reduced superheat' ($s$) [8,9,13,19,21]. Here the data for the different liquids do not follow a single line. In Fig.7, we replot the data used in Fig.6 in terms of the 'effective recoil nucleus threshold energy' with the 'reduced superheat' ($s$). It is observed in this case that all the experimental data for the different liquids fall on a single line. In the plot, $C_4F_{10}$ data is obtained from refs. [19], R114 data from refs. [8,9,13,21], $C_4F_8$ data from refs. [8] and R142B data from refs. [21]. Experimental observation shows that about 90% of the critical temperature can be reached for the superheated liquid droplets [22]. Therefore, we define here, $s'$ by replacing $T_c$ with $0.9T_c$ in the expression of '$s$' and another set of data points with $s'$ is obtained which is shown in Fig.8. In both the cases, either with '$s$' or $s'$, the data points fall on the single curve. Thus, a plot of the 'effective recoil nucleus threshold energy' versus the 'reduced superheat' ($s$ or $s'$) can provide a universal energy calibration curve for superheated droplet detector.

Fig.9 and Fig.10 show the variation of nucleation parameter for the two liquids $C_4F_{10}$ and R114 as a function of effective recoil nucleus threshold energy and the reduced superheat respectively. The solid line is obtained from Eq.(1) and by converting '$k$' to '$a$' using the Eq.(2). The nucleation parameter varies significantly for $C_4F_{10}$ in Fig.9 and in Fig.10 but not as for R114. The possible explanation of Fig.9 and Fig.10 is as follows : the nucleation parameter depends on the energy deposition, and the temperature, pressure of the sensitive liquid. The energy deposition again depends on the mass of the recoil nuclei therefore it depends on the composition of the sensitive liquids. For R114, the effective recoil nucleus is chlorine and for $C_4F_{10}$, it is the fluorine, in the energy range of interest. Fig.4 shows that the $\frac{dE}{dx}$ of chlorine does not vary significantly as it is varied for the case of fluorine in $C_4F_{10}$. Therefore, in Fig.9 and in Fig.10, the results for different liquids do not appear to be on single line while plotted with the effective recoil nucleus threshold energy and the reduced superheat independently and appears to be dependent on both the energy of the particle and the type of the sensitive liquid.

## 4. Conclusions

A comparative study is performed on the response of R114 and $C_4F_{10}$ detector, in a mixed neutron – γ radiation field. Neutron - γ discrimination is observed to be more prominent in R114, the liquid having recoil nuclei with larger values of LET, than that of $C_4F_{10}$. Although the probability of γ induced nucleation increases with increase in temperature in $C_4F_{10}$ liquid droplets, the peak of the power distribution is observed to be independent of the operating temperature. Studies on the nucleation parameter for the two liquids do not provide a universal behaviour while plotted either with reduced superheat or with effective recoil nucleus threshold energy. The present study provides the important information on the energy calibration of such detector. The universal energy calibration curve for superheated droplet detector has been explored successfully utilizing the 'effective recoil nucleus threshold energy' and the 'reduced superheat' that can be utilized both for the neutron detection and the WIMPs dark matter search experiments.

## Acknowledgment

The authors sincerely acknowledge Prof. Viktor Zacek, University of Montreal, Canada and Prof. Pijushpani Bhattacharjee, Saha Institute of Nuclear Physics, Kolkata, India for encouragement and useful discussions. Authors are grateful to Prof. B K Chatterjee, Department of Physics, Bose Institute, Kolkata, India, for providing $^{137}$Cs gamma source.

References

1. R. E. Apfel, US Patent 4143274 (1979).

2. H. Ing and H. C. Birnboim, Nucl. Tracks Radiat. Meas. **8,** 285 (1984).

3. M. Barnabe-Heider, *et al*, PICASSO collaboration, Nucl. Instrum. Meths. A **555**, 184 (2005).

4. T. Morlat, M. Felizardo, F. Giuliani, T. A. Girard, G. Waysand, R. F. Payne, H. S. Miley, A. R. Ramos, J. G. Marques, R. C. Martins, D. Limagne, Astropart. Phys. **30**, 159 (2008).

5. W. J. Bolte, J. I. Collar, M. Crisler, J. Hall, D. Holmgren, D. Nakazawa, B. Odom, K. O'Sullivan, R. Plunkett, E. Ramberg, A. Raskin, A. Sonnenschein, J. D. Vieira, Nucl. Instr. Meth. A **577**, 569 (2007).

6. R. E. Apfel and S. C. Roy, Nucl. Instrum. Meths. **219**, 582 (1984).

7. R. E. Apfel and S. C. Roy, Rad. Prot. Dosim. **10**, 327 (1985).

8. F. D'Errico, Nucl. Instrum. Meths. B **184**, 229 (2001).

9. Mala Das, B. K. Chatterjee, B. Roy, S. C. Roy, Rad. Phys. Chem. **66**, 323 (2003).

10. Mala Das, S. Seth, S. Saha, S. Bhattacharya, P. Bhattacharjee, Nucl. Instrum. Meths. A **622**, 196 (2010).

11. F. Aubin *et. al*., PICASSO collaboration, New J. Phys. **10**, 103017 (2008).

12. S. Archambault *et. al.*, PICASSO collaboration, Phy. Lett. B **682**, 185 (2009).

13. Mala Das, Teroku Sawamura, Nucl. Instrum. Meths. A **531**, 577 (2004).

14. S. Archambault *et.al.*, PICASSO collaboration, New J. Phys. **13**, 043006 (2011).

15. B. Roy, B. K. Chatterjee, S. C. Roy, Radiat. Meas. **29**, 173 (1998).

16. Mala Das, A. S. Arya, C. Marick, D. Kanjilal, S. Saha, Rev. Sci. Instrum. **79(11)**, 113301 (2008).

17. J. F. Ziegler, M. D. Ziegler, J. P. Biersack, SRIM 2008 ( SRIM.com).

18. J. W. Gibbs, Transl. Conn. Acad. III, 108 (1875).

19. PICASSO, University of Montreal group (2010) (personal communications).

20. R. E. Apfel, Y. C. Lo, S. C. Roy, Phys. Rev A **31**, 3194 (1985).

21. S. C. Roy, R. E. Apfel, Y. C. Lo, Nucl. Instrum. Meths. A **255**, 199 (1987).

22. Mala Das, B. K. Chatterjee, B. Roy, S. C. Roy, Phys. Rev E **62**, 5843 (2000).

Figure Captions

**Fig.1.** Observed response of $C_4F_{10}$ detector as a function of temperature in presence of $^{137}Cs$ γ - source.

**Fig.2**. Observed differential power distribution of pulses of $C_4F_{10}$ detector in presence of $^{137}Cs$ γ - source.

**Fig.3**. Variation of threshold neutron energy ($E_{n,th}$) with temperature for $C_4F_{10}$ and R114.

**Fig.4**. Variation of $\frac{dE}{dx}$ for C, Cl, F in R114 and for C, F in $C_4F_{10}$.

**Fig.5**. Observed differential power distributions at different temperatures for $C_4F_{10}$ and R114 in presence of $^{252}Cf$ fission neutron source.

**Fig.6**. Variation of threshold neutron energy ($E_{n,th}$) as a function of reduced superheat, $s$ for different liquids.

**Fig.7**. Variation of effective recoil nucleus threshold energy ($E_{Reff,th}$) as a function of reduced superheat, $s$.

**Fig.8**. Variation of effective recoil nucleus threshold energy ($E_{Reff,th}$) as a function of reduced superheat, $s'$.

**Fig.9**. Nucleation parameter (*a*) as a function of effective recoil nucleus threshold ($E_{Reff,th}$) energy.

**Fig.10**. Nucleation parameter (*a*) as a function of reduced superheat, $s$.

**Table-1**. Physical and calculated parameters for $C_4F_{10}$ and R114.

| Liquids | Density (1.013 bar at b.p.) (gm/cc) | Critical radius at 25°C ($r_c$) (μm) (calculated) | Critical energy at 25°C (W) (keV) (calculated) | Critical temperature (°C) | Critical pressure (bar) |
|---|---|---|---|---|---|
| $C_4F_{10}$ (b.p. -1.7°C) | 1.594 | 0.133 | 4.91 | 113.0 | 24.27 |
| R114 ($C_2Cl_2F_4$; b.p. 3.7°C) | 1.527 | 0.194 | 10.81 | 145.7 | 32.63 |

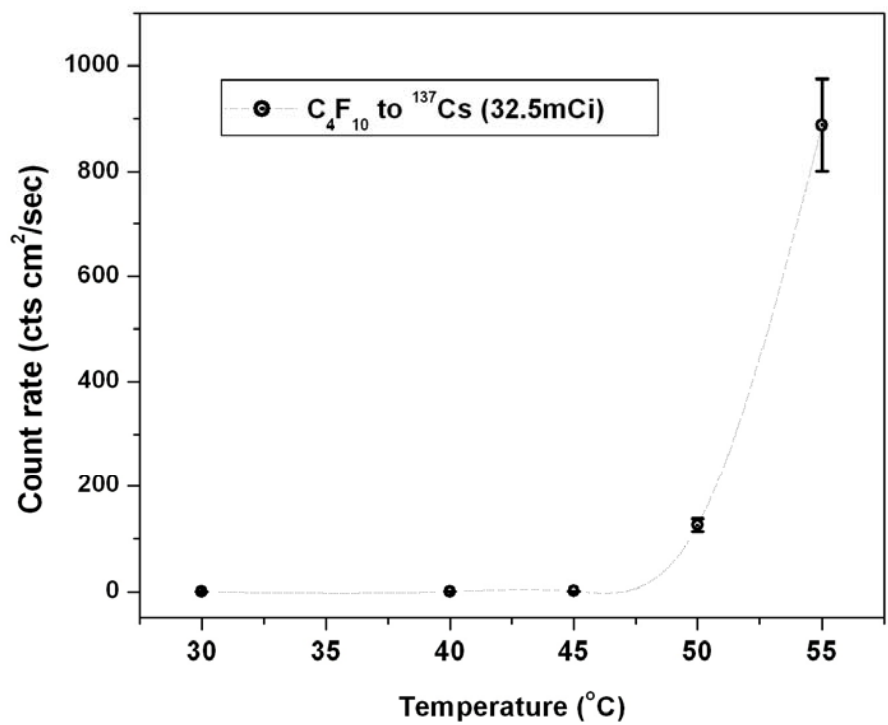

**Fig.1.** Observed response of $C_4F_{10}$ detector as a function of temperature in presence of $^{137}Cs$ γ - source.

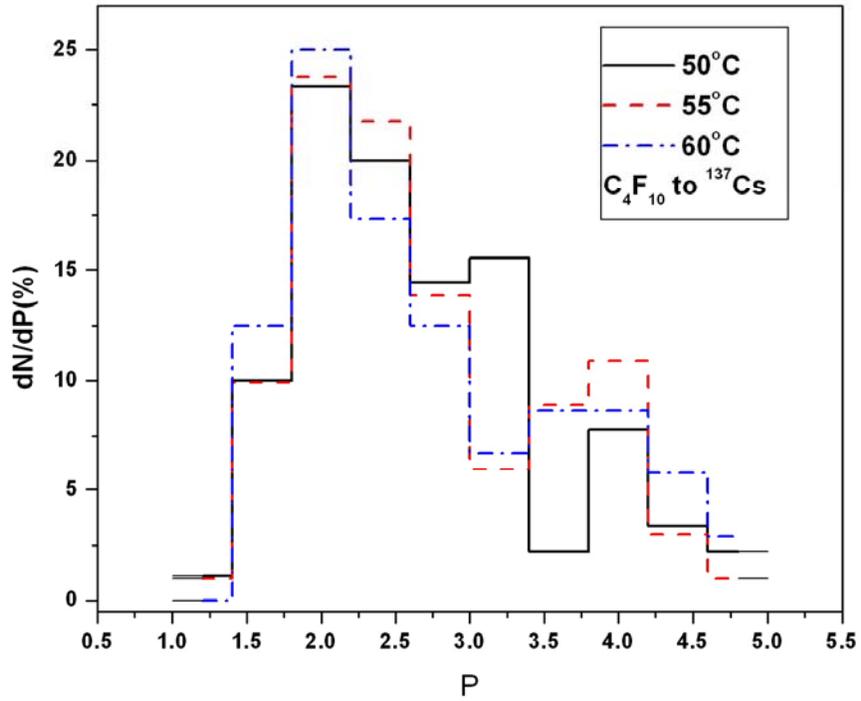

**Fig.2**. Observed differential power distribution of pulses of $C_4F_{10}$ detector in presence of $^{137}Cs$ γ - source.

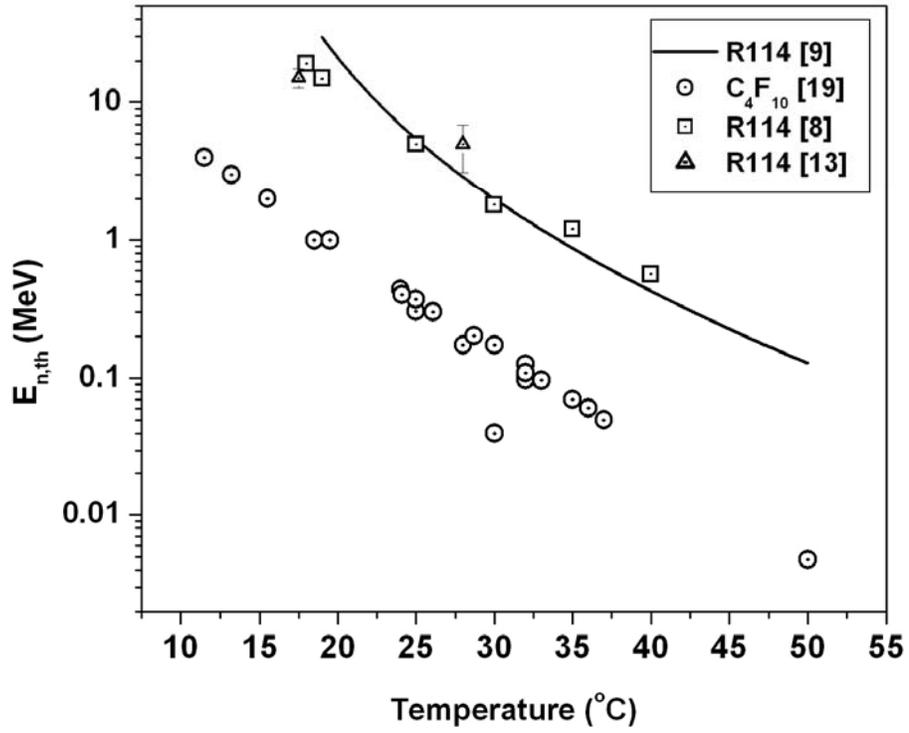

**Fig.3**. Variation of threshold neutron energy ($E_{n,th}$) with temperature for $C_4F_{10}$ and R114.

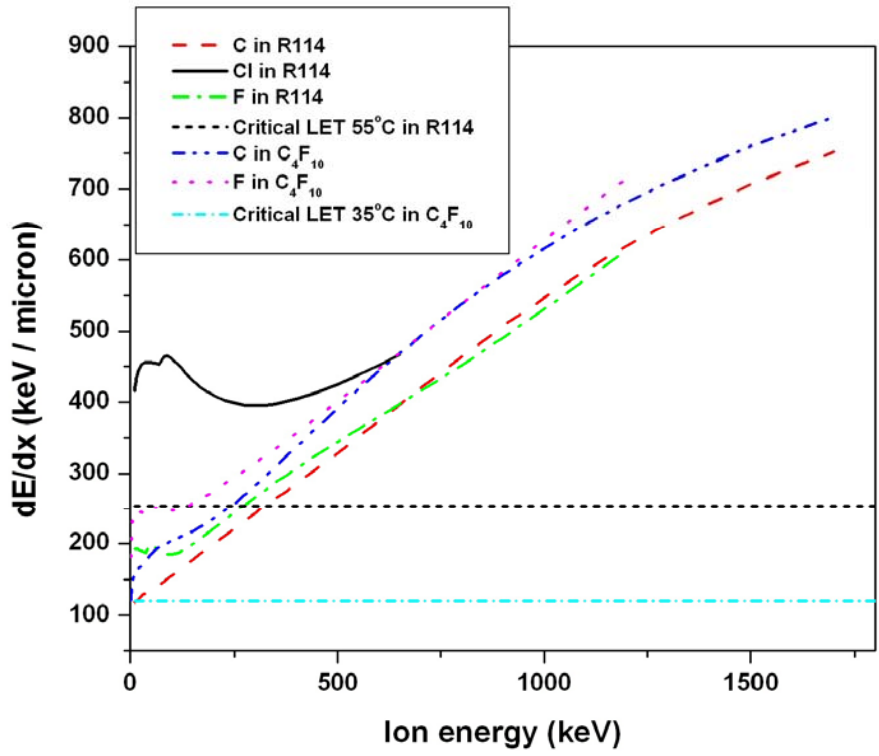

**Fig.4**. Variation of $\dfrac{dE}{dx}$ for C, Cl, F in R114 and for C, F in $C_4F_{10}$.

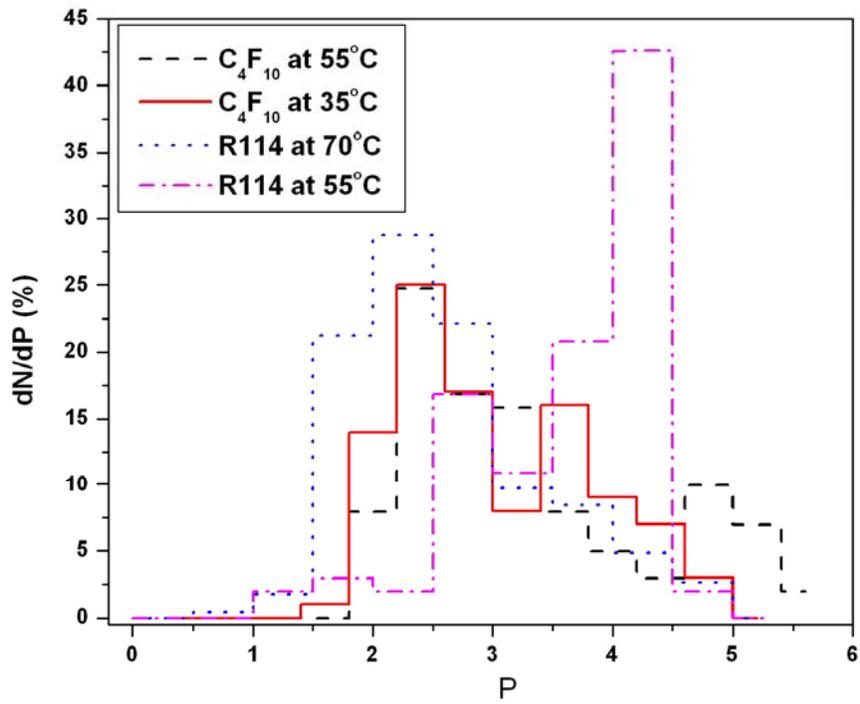

**Fig.5**. Observed differential power distributions at different temperatures for $C_4F_{10}$ and R114 in presence of $^{252}Cf$ fission neutron source.

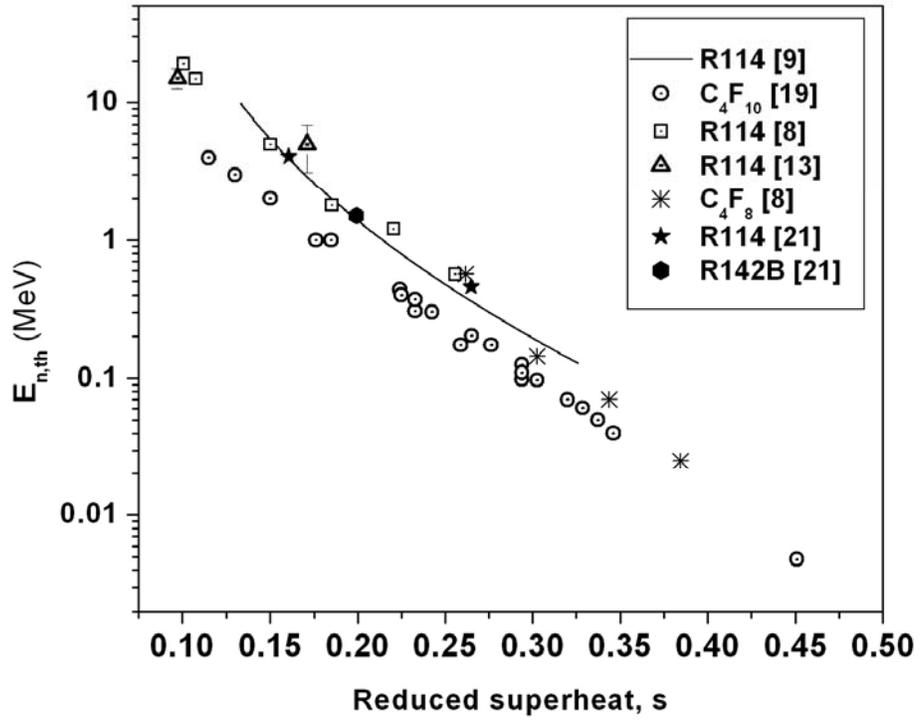

**Fig.6**. Variation of threshold neutron energy ($E_{n,th}$) as a function of reduced superheat, *s* for different liquids.

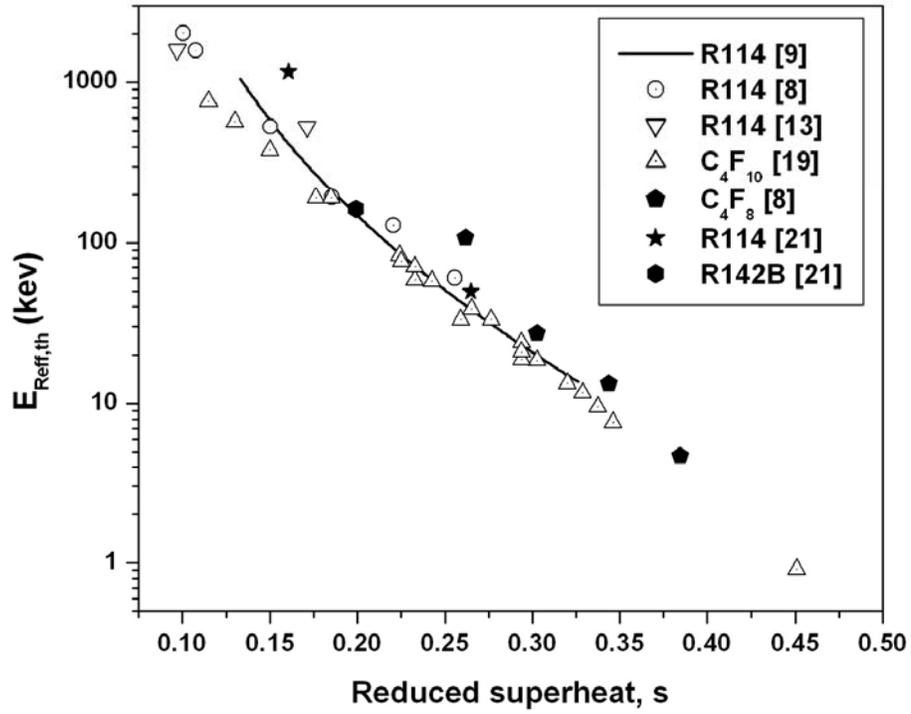

**Fig.7**. Variation of effective recoil nucleus threshold energy ($E_{Reff,th}$) as a function of reduced superheat, *s*.

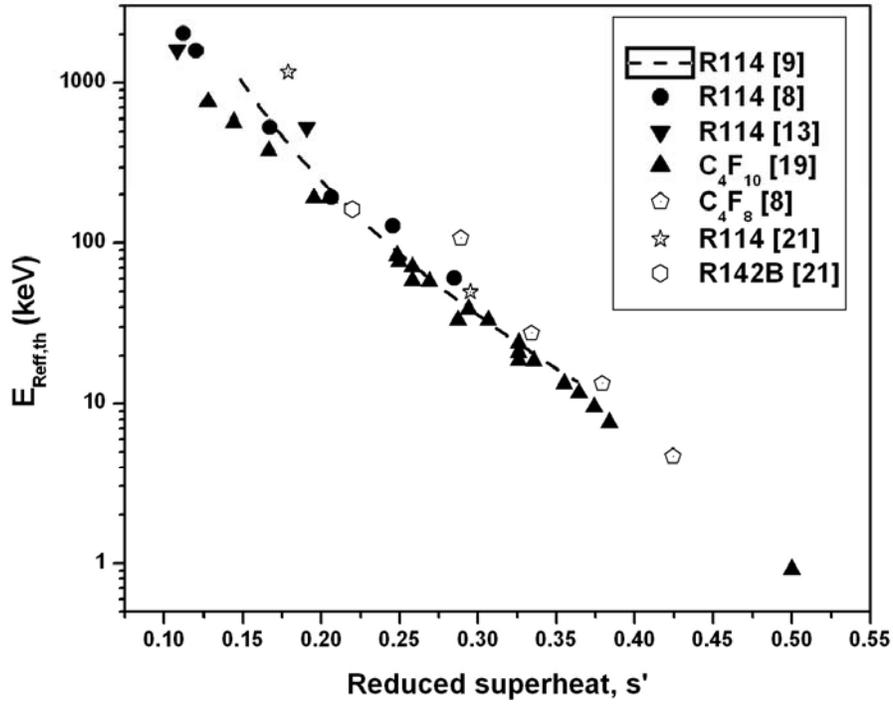

**Fig.8**. Variation of effective recoil nucleus threshold energy ($E_{Reff,th}$) as a function of reduced superheat, $s'$.

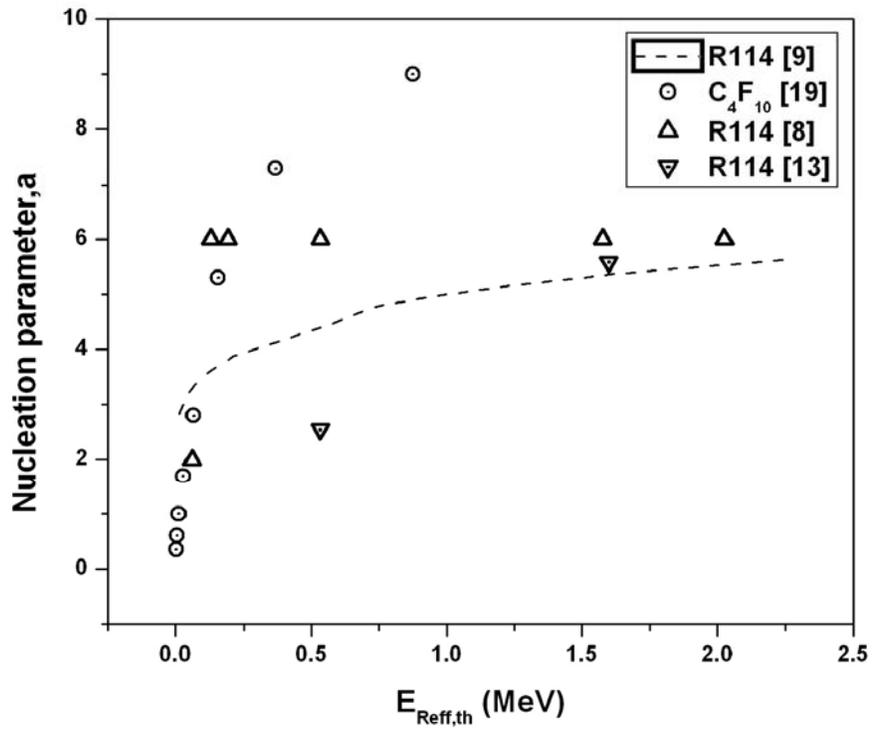

**Fig.9**. Nucleation parameter (*a*) as a function of effective recoil nucleus threshold ($E_{Reff,th}$) energy.

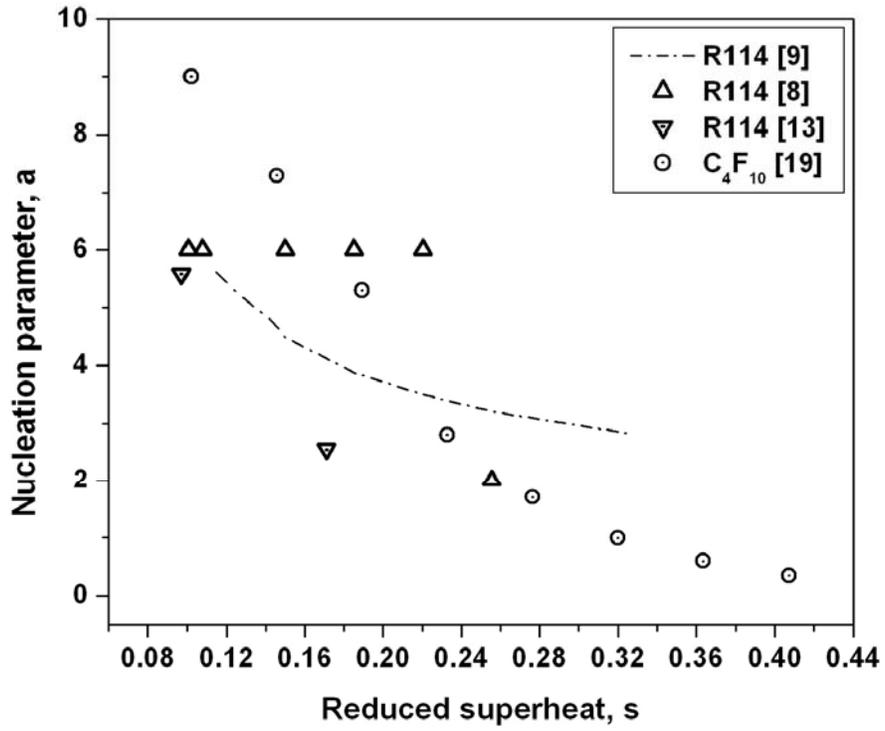

**Fig.10**. Nucleation parameter (*a*) as a function of reduced superheat, *s*.